\documentclass[pra,twocolumn,showpacs,preprintnumbers]
{revtex4}

\usepackage{bm}
\usepackage{graphicx}
\usepackage{amsbsy}
\usepackage{amsmath}
\usepackage{amsfonts}
\usepackage{amsthm}

\begin{document}

\theoremstyle{plain}
\newtheorem{theorem}{Theorem}
\newtheorem{lemma}[theorem]{Lemma}
\newtheorem{corollary}[theorem]{Corollary}
\newtheorem{proposition}[theorem]{Proposition}
\newtheorem{conjecture}[theorem]{Conjecture}

\theoremstyle{definition}
\newtheorem{definition}{Definition}

\theoremstyle{remark}
\newtheorem*{remark}{Remark}
\newtheorem{example}{Example}

\title{No partial erasure of quantum information}
\author {Arun K. Pati}
\email{akpati@iopb.res.in}
\affiliation{Institute of hysics, Sainik School Post, 
Bhubaneswar-751005, Orissa, India}
\author{Barry C. Sanders}
\email{bsanders@qis.ucalgary.ca}
\affiliation{Institute for Quantm Information Science, 
University of Calgary, Alberta T2N 1N4, Canada}
\affiliation{Australian Centre of Excellence for Quantum Computer Technology,
	Macquarie University, Sydney, New South Wales 2109, Australia}

\date{\today}

\begin{abstract}
In complete erasure any arbitrary pure quantum state is transformed to a
fixed pure state by irreversible operation. Here we ask if the
process of partial erasure of quantum information is possible by 
general quantum operations, where partial erasure 
refers to reducing the dimension of the parameter space
that specifies the quantum state. 
Here we prove that quantum information stored in qubits and qudits cannot be
partially erased, even by irreversible operations.
The no-flipping theorem, which
rules out the existence of a universal NOT gate for an arbitrary qubit, 
emerges as a corollary to 
this theorem. The `no partial erasure' theorem is shown to apply to spin 
and bosonic coherent states, with the latter result showing that the 
`no partial erasure' theorem applies to continuous variable quantum 
information schemes as well. The no partial erasure theorem suggests an 
integrity principle that quantum information is indivisible.

\end{abstract}

 
\maketitle

\section{Introduction}

Classical information can be stored in distinct macroscopic states of a
physical system and processed according to classical laws of physics.
That `information is physical' is exemplified by the fact that  
erasure of classical information is an irreversible operation with a 
cost of $k T \log 2 $ of energy per bit operating at a 
temperature~$T$~\cite{Lan61}, which  is
a fundamental source of heat for standard computation~\cite{Ben82}. 
This is the Landauer erasure principle. In quantum
information processing, a qubit cannot be erased by a unitary
transformation (see Appendix A) and is subject to Landauer's principle.  In
recent years considerable effort has been directed toward
investigating possible and impossible operations in quantum
information theory.  

Impossible operations are stated as no-go theorems,
which establish limits to quantum information capabilities and also
provide intuition to enable further advances  in the field. For
example the no-cloning theorem~\cite{Woo82,dd82,yuen86} underscored 
the need for
quantum error  correction to ensure that quantum information
processing is possible in faulty systems  despite the impossibility of
a quantum FANOUT operation. Other examples of important no-go theorems
are the no-deletion theorem~\cite{pb,Zur00}, which proves the
impossibility of perfectly  deleting  one state from two identical
states, the no-flipping theorem~\cite{Buz99}, which  establishes the
impossibility of designing a universal NOT gate for arbitrary qubit
input states, and the  impossibility of universal Hadamard and CNOT
gates for arbitrary qubit input  states~\cite{Pat02}.  The strong
no-cloning theorem states that the creation of a copy of a quantum
state requires full information about the quantum state~\cite{jozsa};
together with the no-deletion theorem, these establish
permanence of quantum information.  A profound consequence of the
no-cloning and no-deleting theorems suggest a fundamental principle of
conservation of quantum information~\cite{Hor03}.

Here we establish a new and powerful no-go theorem of quantum
information,  which suggests both a limitation and protection of
quantum information. Our theorem shows that it is impossible to
erase quantum information, even partially and even by using
irreversible means, where partial erasure corresponds to a reduction
of the parameter space dimension for the quantum state that holds
the quantum information, namely the qubit or qudit. As a special
case, it is impossible to erase azimuthal angle information of a
qubit whilst keeping the polar angle information intact, which we
show is the no-flipping principle. Our theorem adds new insight into
the integrity of quantum information, namely that we can erase 
complete information but not partial information.
This in turn implies that quantum
information is indivisible and we have to treat quantum information as a
`whole entity'. We also introduce the no partial erasure
propositions for SU(2) coherent states and for continuous  variable
quantum information.
Since the first e-print release of our work, a no-splitting theorem  
for quantum information~\cite{Zho05} has been presented, which we show 
is a straightforward 
corollary to our Theorem~\ref{theorem:arbitrary}.

An arbitrary qubit is expressed as 
\begin{equation}
	|\Omega\rangle=\cos\tfrac{\theta}{2} |0\rangle
		+e^{i\phi}\sin\tfrac{\theta}{2}|1\rangle \in {\cal H}^2
\end{equation}
with 
$\Omega\equiv(\theta,\phi)$ and 
\begin{equation}
	\theta\in [0,\pi], \; \phi\in [0,2\pi).
\end{equation}
Each pure state is uniquely identified with a point on the
Poincar\'e sphere with~$\theta$ the polar angle and~$\phi$ the
azimuthal angle.  The states $|0\rangle$ and $|1\rangle$ are the
logical zero and one states, respectively.  Complete erasure would map
all arbitrary qubit states into a fixed qubit state~$|\Omega_0\rangle
= |\Sigma \rangle$  regardless of the input state parameters $\theta$
and $\phi$, which is known to be impossible by unitary means. 
More generally, the $d$-dimensional analogue of the $2$-dimensional 
qubit is a qudit with quantum state
\begin{equation}
\label{eq:qudit}
	|\vec{\Omega}\rangle=\sum_{i=1}^d e^{i \phi_i}
		\cos\tfrac{\theta_i}{2}|i\rangle \in {\cal H}^d,
\end{equation}
with 
\begin{equation}
	\vec{\Omega}\equiv(\vec{\theta},\vec{\phi}),
\end{equation}
and
$\cos\tfrac{\theta_i}{2} = |\langle i |\vec{\Omega}\rangle|$ with $\theta_i$ 
is the Bargmann angle between the
$i^\text{th}$ orthonormal vector and the qudit state.

Each vector $\vec{\theta}$ and $\vec{\phi}$ is $d$-dimensional, but  
normalization
of the qudit state and the unphysical nature of the overall phase
reduces  the number of free parameters for the qudit to $2(d-1)$. A
pure qudit can be  represented as a point on the projective Hilbert
space ${\cal P}$ which is a real $2(d-1)$-dimensional manifold.  For
the qubit case, $d=2$, and there are two parameters, so we see the
reduction of the formula for two qubits is correct.

The organization of our paper is as follows. In the section II, we present
our no-partial erasure theorem for non-orthogonal qubits and qudits. 
Also we show how the no-flipping theorem for arbitrary qubits emerges 
as a corollary to our theorem. Then, we prove the no-partial erasure 
theorem for an arbitrary qudit using linearity of quantum theory. Furthermore, 
we show that the no-splitting theorem for quantum information also follows 
from our theorem.
In the section III, we present the no-partial erasure result for spin 
coherent state. In the section IV, we generalize the no-partial erasure 
theorem for continuous variable quantum information. Lastly, in the section V, 
we conclude our paper.

\section{No-Partial Erasure of Qubit and Qudit}

In this section, we prove powerful theorems that establish the
impossibility of partial  erasure of arbitrary qudit states but first
begin with a definition of partial erasure.
\begin{definition}
Partial erasure is a completely positive (CP), trace preserving
mapping of all  {\em pure} states 
\begin{equation}
	|\{\zeta_i;i=1,\ldots,n\}\rangle,
\end{equation}
with real parameters  $\zeta_i$, in an $n$-dimensional Hilbert space
to {\em pure} states in an $m$-dimensional Hilbert subspace via  a
constraint
\begin{equation}
	\kappa(\{\zeta_i;i=1,\ldots,n\}
\end{equation}
such that $m<n$.
\end{definition}

The process of partial erasure reduces the dimension of the parameter
domain and {\em does not leave the state entangled with any other system}.
One may wonder why we emphasize that the process does not entangle with other
system; it is because we want to analyze this process in parallel with the
complete erasure process. 

Recall that, in complete erasure, an arbitrary
pure state of a qubit is mapped to a fixed pure state, i.e., 
$|\Omega \rangle \mapsto |\Sigma \rangle = |0\rangle$. 
If we allow the original system to
be entangled with ancilla, then we would trivially be able to
erase partial information. 
For example, if we commence with a qubit in the state
\begin{equation}
	|\Omega \rangle = \alpha |0 \rangle + \beta |1 \rangle
\end{equation}
and enjoin an 
ancilla in the state~$|0 \rangle$, then a controlled-NOT gate would 
entangle these two qubits together. 
Then the resulting state of the original qubit has no phase information 
about the input state. So we have a process that maps 
\begin{equation}
	|\Omega \rangle \langle\Omega | \mapsto \rho(\theta).
\end{equation}

Therefore, we do not want that final state
of the quantum system is in a mixed state.
We would like to see if the partial information can
be erased and yet we retain purity of a quantum state in question.
One example of partial erasure would be reducing the parameter space
for the qubit from $\Omega$ to $\theta$ by fixing $\phi$ as a
constant (say $\phi=0)$, i.e.,
\begin{equation}
	|\theta,\phi\rangle \mapsto |\theta \rangle
\end{equation}
corresponds to partial erasure of a qubit where 
the phase information or azimuthal angle information about a qubit is lost.

We now prove that there cannot exist a physical operation capable of
erasing any pair of non-orthogonal qudit states.
\begin{theorem}
\label{theorem:nonorthogonal}
In general, there is no physical operation that can partially erase any
pair of non-orthogonal qudits.
\end{theorem}
\begin{proof} 
The partial erasure quantum operation is a CP,  trace-preserving
mapping that transforms an arbitrary qudit state
$|\vec{\Omega}\rangle$ into the constrained qudit state
$|\vec{\Omega}\rangle_\kappa$ for $\kappa(\vec{\Omega})=0$ a
constraining equation that effectively reduces the parameter space by
at least one dimension.  Arbitrary qudit states can be represented as
points on the projective Hilbert space parametrized by $\Omega$, and
$\kappa$ constrains  these points to a one-dimensional subset of the
projective Hilbert space.

We can introduce the parametrization~$\vec{\tau}$ so that the
constraint~$\kappa$ allows us to write the parameters as
$\vec{\Omega}(\vec{\tau})$.  Consider the mapping of two distinct
qudit states $|\vec{\Omega} \rangle$ and $|\vec{\Omega}' \rangle$ to
$|\vec{\Omega(\tau}) \rangle$ and $|\vec{\Omega}'(\vec{\tau}')
\rangle$, respectively, for some values $\vec{\tau}$ and
$\vec{\tau}'$. 

By attaching an ancilla, the quantum operation ${\cal E}$
that maps
\begin{equation}
	|\vec{\Omega}\rangle \langle \vec{\Omega}| \mapsto 
	\mathcal{E}\left(|\vec{\Omega}\rangle \langle \vec{\Omega}|\right) =
	|\vec{\Omega}(\vec{\tau}) \rangle \langle
	\vec{\Omega}(\vec{\tau})|
\end{equation}
can be represented as a unitary evolution on the enlarged Hilbert
space, so partial erasure of qudits can be expressed as
\begin{eqnarray}
|\vec{\Omega} \rangle |A\rangle &\mapsto & |\vec{\Omega}(\vec{\tau})
	\rangle |A_{\Omega}\rangle,\; \nonumber\\
	|\vec{\Omega'} \rangle |A\rangle & \mapsto &
	 |\vec{\Omega}'(\vec{\tau}')
	\rangle |A_{\Omega'}\rangle,
\end{eqnarray}
where $|A \rangle$ is the initial state, $|A_{\Omega}\rangle$, and 
$|A_{\Omega'}\rangle$  are the final states of 
the ancilla. Now, by unitarity, taking the inner product we have
\begin{equation}
	\langle \vec{\Omega}|\vec{\Omega'}\rangle 
	=\langle  \vec{\Omega}(\vec{\tau})| \vec{\Omega'}(\vec{\tau'})\rangle 
	\langle A_{\Omega}|A_{\Omega'} \rangle.
\end{equation}
However, the inner product of the resultant two qudit states is not
same as the inner product of the original qudit states. Hence we
cannot partially erase a  pair of non-orthogonal qudits by any
physical operation.
\end{proof}

If~$d=2$, we readily obtain the no partial erasure theorem for any pair
of non-orthogonal qubits. 

\begin{example}
As a special, and instructive, case, let us consider the impossibility
of erasing azimuthal angle information for qubits.  Consider the partial
erasure of two non-orthogonal qubit states  $|\Omega \rangle = 
|\theta, \phi \rangle $ and $|\Omega' \rangle = |\theta', \phi' \rangle$ 
by removing the azimuthal angle information.
In the enlarged Hilbert space the unitary transformations for these
two states are given by
\begin{eqnarray}
|\theta, \phi \rangle | A \rangle & \mapsto & 
|\theta \rangle |A_{\Omega}\rangle,  \; \nonumber\\
|\theta', \phi' \rangle | A \rangle
	& \mapsto & |\theta' \rangle |A_{\Omega'}\rangle.
\end{eqnarray}
As unitary evolution must preserve the inner product, we have
\begin{equation}
	\langle \theta, \phi|\theta', \phi' \rangle 
	= \langle \theta|\theta' \rangle \langle A_{\Omega}|A_{\Omega'} \rangle.
\end{equation}

More explicitly, in terms of these real parameters we have
\begin{align}
	\cos \frac{\theta}{2} \cos \frac{\theta'}{2} + 
	\sin \frac{\theta}{2} \cos \frac{\theta'}{2} e^{i (\phi' -\phi)} = 
	\cos \frac{\theta - \theta'}{2} \langle A_{\Omega}|A_{\Omega'} \rangle.
\end{align}
However, for arbitrary values of $\phi$ and $\phi'$ the above equation
cannot hold.  Therefore, it is impossible to erase azimuthal angle
information of a qubit by physical operations.
\end{example}

The above equation suggests that there may be special classes of qubit
states that can be partially erased. The general condition is  that if
\begin{equation}
	\phi = \phi' + 2n\pi,
\end{equation}
$n$ being an integer, then any qubit that
differs in phase by $2n\pi$ can be partially erased. This implies if
we restrict our  qubits to be chosen from any great circle passing
through north and  south poles of the Poincar\'e sphere, then those
qubits can be partially erased by a physical operation. Similarly, we
can show that  it is impossible to erase the information about the
polar angle $\theta$ of an arbitrary qubit, i.e., the transformation
\begin{equation}
	| \theta, \phi \rangle | A \rangle \mapsto | \phi \rangle |A_{\Omega}\rangle
\end{equation}
is not allowed.

\begin{remark}
	Although there does not exist a completely positive, trace-preserving mapping
	that  partially erases a qubit, there exists a proper subset of qudit or 
	qubit states that are erased by a given mapping. For example the set of
	qubit states whose parameters satisfy the  constraint~$\kappa$ can
	have partial erasure according to the already-imposed
	constraint~$\kappa$. Partial erasure can also be effected on an
	arbitrary qubit by a unitary  mapping if the state is known simply
	because there always exists a unitary map between any two  states in
	 a Hilbert space; hence there exists a unitary mapping from every
	qubit  state to  constrained qubit states. Also a qubit or a qudit 
	in known orthogonal states can  be partially erased.
\end{remark}

Now we show that for $d=2$ and
\begin{equation}
	\kappa(\Omega=(\theta,\phi))= \kappa(\theta,\phi_0)
\end{equation}
for all $\Omega$, with the azimuthal phase~$\phi_0$ fixed, we obtain the
no flipping principle for an arbitrary qubit. We know that a
classical bit like $0$ or $1$ can be flipped, so also a qubit in an 
orthogonal state like $|0\rangle$ or $|1\rangle$. However, an unknown qubit 
$|\Omega \rangle$ cannot be flipped. That is there is no exact universal NOT 
gate for an arbitrary qubit. This is because the flipping operation is 
an anti-unitary operation which is not a CP map and thus cannot be implemented
physically. The no-flipping principle
for an unknown qubit is another important limitation in quantum information.

\begin{corollary}
Erasure of the azimuthal phase from the parameter domain of a qubit,
whilst leaving the polar phase parameter unchanged by the mapping,
implies the existence of a universal NOT gate.
\label{lemma:polarerasure}
\end{corollary}
\begin{proof} Suppose we can erase phase information of an arbitrary qubit.
For an orthogonal qubit state $|\theta, \phi\rangle^\perp$
partial erasure effects the mapping 
\begin{equation}
	|\theta, \phi \rangle^\perp|A\rangle\mapsto|\theta \rangle^\perp |A_{\Omega}\rangle^\perp.
\end{equation}
If this holds, then after the partial erasure one can apply a local
unitary NOT gate to $|\theta\rangle^\perp$ and convert it to
$|\theta\rangle $ (in this case by applying $i \sigma_y$). 

Next an application of  the inverse of the partial erasure transformation
yields the state $|\theta,\phi \rangle$.  This means by applying a
sequence of unitary transformations one can flip an unknown qubit
state, that is, map an arbitrary qubit to its complement. Hence
erasure of  azimuthal phase but not polar phase implies the existence
of a universal NOT gate.
\end{proof}

Now we can apply the above to prove easily the
non-existence of a universal NOT gate~\cite{Buz99}.
\begin{corollary}
	A universal NOT gate is impossible.
\end{corollary}
\begin{proof} To prove the no flipping principle, we show that a universal
NOT gate requires partial erasure. Suppose there is a universal NOT
gate for an arbitrary qubit that takes
\begin{equation}
	|\theta, \phi\rangle \mapsto |\theta, \phi \rangle^\perp.
\end{equation}
However, it is known that such an
operation exists~\cite{ghosh,arun} if and only if the qubit belongs to
a great circle, that is, the qubit  parameter domain is constrained by
$\kappa$ to a great circle on the sphere defined by $\theta$ and
$\phi$. This means that the arbitrary qubit must have been mapped to a
qubit on the great circle (this mapping is a partial erasure machine)
before passing through the universal NOT.

After the universal NOT it must have passed through a reverse partial
erasure machine. Thus to be able to design a universal NOT gate for an
arbitrary qubit we need a partial erasure operation from
\begin{equation}
	|\theta,\phi \rangle \mapsto |\theta \rangle.
\end{equation}
However, we know that this
is impossible. Hence no partial erasure of phase information implies the 
non-existence of a universal NOT gate for a qubit.
\end{proof}

\begin{remark}
Theorem~\ref{theorem:nonorthogonal} establishes that there is no
physical operation that can partially erase any pair of nonorthogonal
qudit states, from which the `no flipping principle' emerges as a
simple corollary. However, Theorem 1 applies to a set of quantum states 
which are not arbitrary. One can ask a more general question: Can we 
partially erase an arbitrary quantum state by a linear transformation?
Now we show that linearity of quantum theory
establishes that there cannot exist a physical operation that can
partially erase a qudit, which is a stronger result.
\end{remark}

\begin{theorem} 
\label{theorem:arbitrary}
	An arbitrary qudit cannot be partially erased by an
	irreversible operation.
\end{theorem}
\begin{proof} We know that partial erasure 
operation for known orthogonal states is possible. Let
$|\vec{\Omega}_n \rangle$ be a known orthonormal basis in 
$\mathcal{H}^{d}$. 
Then a partial erasure operation for these states  yields
\begin{equation}
	|\vec{\Omega}_n \rangle | A \rangle \mapsto
	|\vec{\Omega}_n(\vec{\tau}) \rangle  |A_{\Omega_n}\rangle.
\end{equation}
Consider an arbitrary qudit $|\vec{\Omega} \rangle$ of 
Eq.~(\ref{eq:qudit}) which is a
linear superposition of the basis states $\{ |\Omega_n \rangle \}$.
Suppose partial erasure of  $|\vec{\Omega} \rangle$ is possible.  Then
linearity of the partial erasure transformation requires that
\begin{align}
	|\vec{\Omega} \rangle  | A \rangle  &= \sum_{n=1}^d e^{i \phi_n}
	\cos\tfrac{\theta_n}{2}|\vec{\Omega}_n \rangle |A \rangle \mapsto 
		\nonumber\\
	& \sum_{n=1}^d e^{i \phi_n}
	\cos\tfrac{\theta_n}{2}|\vec{\Omega}_n(\vec{\tau}) \rangle 
	|A_{\Omega_n}\rangle
	=| \bf{\Omega} \rangle.
\end{align}
Ideally the partial erasure of an
arbitrary qudit should have yielded a {\em pure} output state that takes
constrained values for $\theta_n$ and $\phi_n$.  However, the resultant
state is not a pure qudit state but rather is
entangled with the ancilla.
By definition partial erasure maps a pure state to a pure state, hence a 
contradiction. Thus, linearity (including irreversible operations) prohibits
partial erasure of arbitrary quantum information.  
\end{proof}
For $d=2$ we obtain
the impossibility of partial erasure of an  arbitrary qubit. For example, 
we cannot omit either polar or azimuthal angle information of a qubit 
by irreversible operation.

Here we give another proof for no-partial erasure of an arbitrary qubit solely using
linearity and without using ancilla states.
Suppose we have the partial erasure operation for
two known orthogonal states such as $| \Psi(\theta_0, \phi_0) \rangle$ and 
$| {\bar \Psi}(\theta_0, \phi_0) \rangle$. Then the partial erasure operation 
can be represented by 
\begin{align}
	&| \Psi(\theta_0, \phi_0) \rangle \mapsto | \psi(\theta_0) \rangle,
	 \nonumber\\
	&| {\bar \Psi}(\theta_0, \phi_0) \rangle
	\mapsto | {\bar \psi}(\theta_0) \rangle.
\end{align}

Let an arbitrary qubit $| \Phi(\theta, \phi) \rangle$ be in a linear 
superposition of these two basis states:
\begin{equation}
| \Phi(\theta, \phi) \rangle =
\cos \frac{\theta}{2} | \Psi(\theta_0, \phi_0) \rangle + 
\sin \frac{\theta}{2} \exp(i \phi) | {\bar \Psi}(\theta_0, \phi_0) \rangle .
\end{equation}
If we want to have partial erasure of $| \Phi(\theta, \phi) \rangle$ then,
by linearity of the erasure transformation we have 
\begin{align}
| \Phi(\theta, \phi) \rangle = &
\cos \frac{\theta}{2} | \Psi(\theta_0, \phi_0) \rangle + 
\sin \frac{\theta}{2} e^{i \phi} | {\bar \Psi}(\theta_0, \phi_0) \rangle
\nonumber\\
&\mapsto
\cos \frac{\theta}{2} | \psi(\theta_0) \rangle + 
\sin \frac{\theta}{2} e^{i \phi} | {\bar \psi}(\theta_0) \rangle  \nonumber\\
=&
| {\tilde \Phi}(\theta, \phi) \rangle
\end{align}
Again, ideally the partial erasure of an arbitrary qubit should have 
yielded an output state that is completely independent of $\phi$, i.e., 
$| \Phi(\theta, \phi) \rangle \mapsto | \Phi(\theta) \rangle$.
However, we have a state 
$| {\tilde \Phi}(\theta, \phi) \rangle$ that has complete information about 
both $\theta$ and $\phi$. Hence, this shows that 
linearity (which includes also irreversible operations) 
of quantum theory does not allow partial erasure of quantum information. 
If we include ancilla, then the original qubit will be entangled with the
ancilla and by throwing out ancilla, we will be left with a qubit state 
that is no more pure. Note that if we allow 
irreversible operation (unitary evolution of combined system and tracing 
out of the ancilla), we can eliminate complete information of an
arbitrary qubit (albeit the fact that the original information still remains
in the ancilla) as in the complete erasure. Thus, one can erase the complete 
information of a qubit but not the partial information by an irreversible 
operation and yet retain its purity.

The implication of being able to completely erase, but not partially erase
quantum information implies that quantum information is  
indivisible. There is no classical analogue for this result: 
no partial erasure is a strictly quantum phenomenon.
We introduce the term \emph{integrity principle} to 
refer to this inability to partially erase quantum information.


Since the release of our e-print 
proving `no partial erasure' theorem a `no-splitting theorem' for
quantum information has been presented ~\cite{Zho05},
where `no splitting' refers to the impossibility of splitting 
a qubit~$|\theta,\phi\rangle$ into a product state~
$|\theta\rangle|\phi\rangle$ with one qubit representing the 
$\theta$ information and the other representing
the $\phi$ information. Here we show that the
no-splitting follows from Theorem~\ref{theorem:arbitrary}.
\begin{corollary}
\label{corollary:nosplitting}
	No-partial erasure theorem implies a no-splitting of quantum information.
\end{corollary}
\begin{proof}
Suppose quantum information can be split.
Then there exists an operation that transforms
\begin{equation}
	|\theta, \phi \rangle \mapsto |\theta \rangle |\phi \rangle.
\end{equation}
We can append an ancillary qubit in a specific state and swap
with the second qubit, then trace to eliminate all information about~$\phi$.
Thus splitting implies partial erasure, which contradicts
Theorem~\ref{theorem:arbitrary}. Hence, it is impossible to split quantum 
information.
\end{proof} 

\section{No-partial erasure of spin coherent state}

Our theorem that no partial erasure of qudits is possible is important
because  quantum information is clearly not only conserved but also
indivisible.  However, the `no partial erasure' theorem yields another
important result for erasure of spin coherent states, also known as
SU(2)  coherent states~\cite{Are72,gilm72,perel72}.

The SU(2) coherent states are a generalization  of qubits, which can
be thought of as spin-$\tfrac{1}{2}$ states, to states of higher
spin~$j$. The SU(2) raising and lowering operators are  $\hat{J}_+$
and $\hat{J}_-$, respectively, and their commutator
\begin{equation}
	[\hat{J}_+,\hat{J}_-]=2\hat{J}_z
\end{equation}
yields the weight operator
$\hat{J}_z$  with spectrum 
\begin{equation*}
	\{m;-j \leq m \leq j\}
\end{equation*}
and integer spacing between  successive values of~$m$. The weight basis is
$|j\,m\rangle$ with $j(j+1)$  the eigenvalue for states in the
$j^\text{th}$ irrep of the Casimir invariant  $\hat{J^2}$.

The SU(2) coherent states are obtained by `rotations' of the
highest-weight  state~$|j\,j\rangle$. Here we use the stereographic
parameter
\begin{equation}
	\gamma =  e^{i\phi} \tan(\theta/2)
\end{equation}
that corresponds to the
coordinates of the state  on the complex plane obtained by a
stereographic projection of the point  on the Poincar\'{e} sphere for
the given state, with parameters $\theta$ and $\phi$ are the polar and
azimuthal angular coordinates of the Poincar\'e  sphere defined
earlier; here the sphere represents states of a  $(2j+1)$-dimensional
system, not just the two-dimensional qubit.  For $|j\,j\rangle$ the
highest-weight state, the SU(2) coherent state  is~\cite{San89}
\begin{equation}
|j,\gamma\rangle=R_j(\gamma)|j\,j\rangle =  \sum_{m=0}^{2j}
	\begin{pmatrix} 2j \\ m \end{pmatrix}^{1/2}
	\frac{\gamma^m}{(1+|\gamma|^2)^j} |j\,j-m\rangle
\end{equation}
for
\begin{align}
	R_j(\gamma)&=\exp\left[\tfrac{1}{2}\theta\left(\hat{J}_-e^{i\phi}-\hat{J}_
	+e^{-i\phi}\right)\right] \nonumber \\ &= \exp(\gamma \hat{J}_-)
	\exp[-\hat{J}_z \ln(1 + |\gamma|^2 )]
	\exp(- \gamma^*\hat{J}_+ ).
\end{align}

We can now prove the following no go result using our theorem.
\begin{corollary}
	Partial erasure of SU(2) coherent states is impossible.
\end{corollary}
\begin{proof}
The SU(2) coherent state is a qudit with the constraint that
\begin{equation}
	e^{i\phi_m}\cos\tfrac{\theta_m}{2}
		=\begin{pmatrix}2j\\m\end{pmatrix}^{1/2}\frac{\gamma^m}{(1+|\gamma|^2)}
\end{equation}
for each~$m$. Partial erasure of the SU(2) coherent states corresponds
to  partial erasure over a subspace of qudits, which we have shown is
impossible.
\end{proof}

\section{No-Partial erasure of Continuous Variable state}

Next, we prove the `no partial erasure'  theorem for continuous
variable quantum information. Ideally  continuous variable (CV) quantum
information encodes quantum information as  superpositions of
eigenstates of the position operator $\hat{x}$, namely
\begin{equation}
	\hat{x}|x\rangle= x|x\rangle; \{ x\in\mathbb{R}\}
\end{equation}
with complex 
amplitude $\Psi(x)$ \cite{qicv}.  We
can represent a CV state as 
\begin{equation}
	|\Psi\rangle=\int_\mathbb{R} dx \Psi(x)|x\rangle,\;\;\Psi(x)=\langle
		 x|\Psi\rangle.
\end{equation}
Note that $\Psi(x)$ can be any complex-valued function, subject to the
requirement of square-integrability and normalization.

Now let us reduce $\Psi(x)$ to a real-valued function, so we have effectively 
reduced the parameter space even in infinity dimensional Hilbert space. 
Does there exist a completely
positive, trace preserving mapping from the set of states with
$\Psi(x)$ a general complex-valued function to the  new $\psi(x)$ a
general real-valued function?

\begin{definition}
Partial erasure of continuous variable quantum information is a
completely positive map of all arbitrary pure states with complex
wavefunctions to pure states with real wavefunctions.
\end{definition}

\begin{theorem}
	There is no physical operation that can partially erase any
	pair of complex wavefunctions.
\end{theorem}
\begin{proof} 
We prove this theorem for a system with one degree of freedom, namely
canonical position~$x$; this proof is readily extended to more than one 
degrees of freedom.  Suppose there is a CP map that can partially erase a
wavefunction $\Psi(x)$ via
\begin{equation}
	|\Psi \rangle |A\rangle \mapsto |\psi \rangle |A_{\Psi}\rangle,\;
\end{equation}
where
\begin{equation}
	||\Psi||^2 = \int_\mathbb{R} dx \; |\Psi(x)|^2,\;\; || \psi||^2 =
	\int_\mathbb{R} dx \; \psi(x)^2.
\end{equation}
If this holds for another arbitrary wavefunction $\Phi(x)$, then we
have
\begin{equation}
	|\Phi \rangle |A\rangle \mapsto |\phi \rangle |A_{\Phi}\rangle,\;
\end{equation}
where
\begin{equation}
	||\Phi||^2 = \int_\mathbb{R} dx \; |\Phi(x)|^2
\end{equation}
and
\begin{equation}
	||\phi||^2 = \int_\mathbb{R} dx \; \phi(x)^2.
\end{equation}
However, the inner product preserving condition
\begin{equation}
	\int_\mathbb{R} dx\, \Psi(x)^*\Phi(x) = \int_\mathbb{R} dx\,\psi(x) \phi(x)
	\int_\mathbb{R} dx\, A_{\Psi}^*(x) A_{\Phi}(x)
\end{equation}
cannot hold for general complex-valued wavefunctions. Hence, we cannot 
partially erase a pair of complex wavefunction.
\end{proof}

This result is analogous to partial erasure of qudits. Furthermore,
the restriction should apply for any erasure of the complex domain by
one dimension (such as a circle where amplitude is fixed and phase
varies).  Similarly, one can give a general proof of no partial
erasure  of continuous variable quantum information, not just complex
to real but  complex to any one-dimensional subset of the complex
space. 

For example, let the partial erasure process transforms the wavefunction
such that one of the complex  amplitudes becomes real (which is one way
to reduce the parameter space  by one dimension). Then we can prove that it
is also impossible.  Under partial erasure the continuous variable state
\begin{equation}
	|\Psi\rangle= \int_\mathbb{R} dx \Psi(x)|x\rangle,
\end{equation}
with
\begin{equation}
	\Psi(x) = \sum_{n=0}^{\infty} c_n \Psi_n(x),
\end{equation}
$c_n = |c_n|\exp(i\theta_n)$  transforms as
\begin{equation}
	\sum_{n=0}^{\infty} c_n \Psi_n(x) \mapsto \sum_{n=0}^{\infty} d_n \Psi_n(x)
\end{equation}
where the constraint is
that all $c_n = d_n$ are complex except for one  $d_k$, which is a real
number. Consider a pair of wavefunctions
$(\Psi(x),\Phi(x))$ 
with
\begin{equation}
	\Phi(x) = \sum_{n=0}^{\infty} c_n' \Phi_n(x),
\end{equation}
$c_n' =  |c_n'| \exp(i\theta_n')$ and 
partial erasure of $\Phi(x)$ is given by
\begin{equation}
	\sum_{n=0}^{\infty} c_n' \Phi_n(x) \mapsto \sum_{n=0}^{\infty} d_n' \Phi_n(x)
\end{equation}
with similar constraints.

For clarity, let us not include an ancilla.  
Unitarity implies that
\begin{equation}
	\exp(i[ \theta_k' -\theta_k]) = 1,
\end{equation}
which is impossible for arbitrary values
of  $\theta_k$ and  $\theta_k'$. Hence, we cannot forget even one
parameter of the complex  wavefunction.

\section{Conclusions}

In summary we have introduced a new process called partial erasure of quantum
information and asked if quantum information can undergo partial erasure.
We have shown that partial erasure of qubits, qudits, SU(2)
coherent states, and continuous variable quantum information is impossible.
These results point to the integrity principle for quantum information,
namely that it is indivisible and robust even against partial erasure.
This principle gives a new meaning to quantum information and 
nicely complements the recent profound observation of
the principle of conservation of quantum information~\cite{Hor03}.

Furthermore, the impossibility theorems presented here underscore
essential differences between classical information (which could be 
stored in 
orthogonal quantum states) and general quantum information,
analogous to related but distinct impossibility results such as the 
no-cloning, no-deleting and no-flipping principles.
Our principle of quantum information integrity may have implications for 
investigations into quantum mechanics over real, complex,
and quaternionic number fields~\cite{Adl95,Stu60}: a unitary equivalence
between complex and real quantum theories would appear to 
contradict the no partial erasure theorem. Interesting problems that 
warrants further investigation is approximate deterministic partial erasure 
and exact probabilistic partial erasure over restricted classes of states.


\vskip 1cm

\emph{Acknowledgments: ---} AKP thanks C. H. Bennett for useful 
discussions. BCS has been supported by Alberta's Informatics Circle 
of Research Excellence (iCORE), the Canadian Institute for Advanced Research,
and the Australian Research Council.

\appendix
\section{No complete erasure by reversible operations}

In quantum theory a reversible operation can be represented by a 
unitary operator. Erasure of a qubit state $|\Psi\rangle$ 
transforms this to a fixed state $|\Sigma \rangle$, which contains no 
information about the input state. 

Consider erasure of a pair of qubits~$|\Omega \rangle$ and $|\Omega' \rangle$ 
such that 
\begin{equation}
	|\Omega \rangle \mapsto |\Sigma \rangle
\end{equation}
and 
\begin{equation}
	|\Omega' \rangle \mapsto |\Sigma \rangle.
\end{equation}
As unitary evolution preserves the inner product, we will have 
\begin{equation}
	\langle  \Omega |\Omega' \rangle = 1,
\end{equation}
which cannot be true. Furthermore, even for two orthogonal states such as 
$|0\rangle$ and $|1\rangle$, this evolution implies a contradiction. 

This paradox demonstrates, in a simplest and yet profound way, that 
neither classical information nor quantum information can be erased by any
reversible operation.


\end{document}